\documentclass[prl,twocolumn]{revtex4}

\usepackage{graphicx,amsmath}

\begin{document}

\title{Alternative Size and Lifetime Measurements for High-Energy Heavy-Ion
Collisions}

\author{Scott Pratt}
\affiliation{Department of Physics and Astronomy, Michigan State
University, East Lansing Michigan, 48824}
\author{Silvio Petriconi}
\affiliation{Department of Physics and Astronomy, Michigan State
University, East Lansing Michigan, 48824}
\date{\today}

\begin{abstract}
\bigskip
Two-Particle correlations based on the interference of identical particles has
provided the chief means for determining the shape and lifetime of sources in
relativistic heavy ion collisions.  Here, Strong and Coulomb induced
correlations are shown to provide equivalent information.

\end{abstract}

\maketitle

\label{sec:intro} 
In the collision of highly relativistic heavy ions, a highly excited region of
matter is created which occupies thousands of fm$^3$ and explodes and dissolves
on a time scale of 10 fm/c. Only the momenta of the collision debris are
experimentally accessible. However, space-time information is crucial for
understanding the reaction \cite{heinzjacak}. Most importantly, a latent heat
associated with a phase transition would be accompanied by a significant
enhancement to the duration of the emitting
phase\cite{prattprd,rischkehydro,teaney,bassdumitrusoff}. The most direct means
for determining the spatial and temporal characteristics of the reaction is
through the experimental measure of the two-particle correlation function which
describes the ratio of the two-particle probability to the product of the
single-particle probabilities for emitting species $a$ and $b$ with momenta
$p_a$ and $p_b$.

The correlation of two particles with outgoing momenta, $p_a$ and $p_b$, is
related to the probability of the source points being separated by a distance
${\bf r}$, $g({\bf r})$, as measured in the two particle center-of-mass frame.
\begin{equation}
\label{eq:corrdef_shorthand}
C({\bf q})=\int d^3r g({\bf r}) |\psi({\bf q},{\bf r})|^2.
\end{equation}
Here, ${\bf q}={\bf p}'_a=-{\bf p}'_b$ is the relative momentum in the
center-of mass frame, and $g({\bf r})$ describes the normalized probability
that the source points of the two particles are separated by a distance ${\bf
r}$. If the particles do not interact with one another, the wave function,
$\psi({\bf q},{\bf r})$, would be $e^{i{\bf q}\cdot{\bf r}}$, and the resulting
correlation would be unity. Since the correlation function can be analyzed for
any total momentum, ${\bf p}_a+{\bf p}_b$, and for any choice of species $a$
and $b$, a wealth of information regarding the space-time characteristics of
the source is available.

A measurement of $C({\bf q})$ represents three dimensions of information and
can lead to a unique extraction of $g({\bf r})$. The temporal information
contained in the four-dimensional source function $S({\bf r},t)$, which
describes the probability of emitting particles separated by ${\bf r}$ and time
$t$ is folded into $g({\bf r})$,
\begin{equation}
g({\bf r})=\int dt S({\bf r}-{\bf v}t,t),
\end{equation}
where ${\bf v}$ is the average velocity of the emitted pair. If the
characteristic time scale of $S({\bf r},t)$ is large, $g({\bf r})$ will be
spread out along the direction of ${\bf v}$ relative to the other
dimensions. In correlation analyses, it is standard to define the direction of
${\bf v}$ as the ``outwards'' direction. The other two dimensions are referred
to as ``longitudinal'' (along the beam axis) and ``sidewards'' (perpendicular
to the beam and to the outwards direction). Assuming that the spatial
dimensions of $S({\bf r},t)$ are roughly equivalent in the outwards and
sidewards direction, one can estimate the lifetime by the difference in the two
dimensions of $g({\bf r})$,
\begin{equation}
v^2\tau^2 \approx R_{\rm out}^2 - R_{\rm side}^2,
\end{equation}
where $R_{\rm out}$ and $R_{\rm side}$ are the spatial dimensions extracted by
a three-dimensional analysis of $C({\bf q})$ with no consideration of lifetime.
This estimate for $\tau$ should work well for large lifetimes.

Remarkably, analyses of two-pion interferometry at RHIC show equal outwards and
sidewards sizes which suggests extremely sudden emission
\cite{starhbt,phenixhbt,phoboshbt}. Dynamical models that incorporate a phase
transition with a latent heat strongly disagree with this observation
\cite{teaney,bassdumitrusoff}. The two-pion analyses are based on the
interference associated with symmetrizing same-sign pions. As the existence or
non-existence of a latent heat with the QCD deconfinement transition represents
a central issues of the RHIC program, it is imperative to explore alternative
means for measuring the size and shape of the emission sources.

For non-interacting identical particles the wave function has a simple form,
\begin{equation}
|\psi({\bf q},{\bf r})|^2=1\pm\cos(2{\bf q}\cdot {\bf r}).
\end{equation}
By performing an inverse Fourier transform of $C({\bf q})-1$ in
Eq. (\ref{eq:corrdef_shorthand}), one can obtain $g({\bf r})$ .  Even though
the pions are charged, experimental analyses have tried to ignore the Coulomb
interaction as much as possible, and in fact try to ``correct'' their data as
to best eliminate the effects of Coulomb from the correlation function so that
comparison with simple forms for $g({\bf r})$ is easily accommodated.  The aim
of this paper is to demonstrate that Coulomb and strong interactions between
the pair should not only be included in correlation analyses, but that they
provide tremendous insight into both the size and shape of $g({\bf r})$.

\begin{figure}[t]
\centerline{\includegraphics[width=0.4\textwidth]{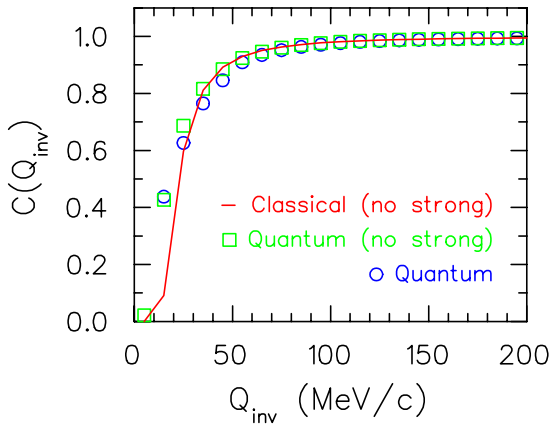}}
\caption{\label{fig:corr_pk} $pK^+$ correlations are shown for a Gaussian
  source ($R_{\rm long}=R_{\rm side}=4$ fm, $R_{\rm out}=8$ fm). The classical
  approximation well explains Coulomb correlations at large relative
  momentum. The strong interaction only moderately affects the correlation
  function.}
\end{figure}

First, we consider two particles which interact only via the Coulomb
interaction. Correlations for $pK^+$ from a Gaussian source,
\begin{equation}
g({\bf r})=\frac{1}{(4\pi)^{3/2}R_{\rm out}R_\perp^2}
\exp\left\{
-\frac{r_{\rm out}^2}{4R_{\rm out}^2}
-\frac{r_{\rm side}^2+r_{\rm long}^2}{4R_{\perp}^2}\right\},
\end{equation}
are displayed in Fig. \ref{fig:corr_pk} as a function of $Q_{\rm inv}=2q$. The
source sizes are chosen $R_{\rm out}=8$ fm and $R_{\perp}=4$ fm. The
integration described in Eq. (\ref{eq:corrdef_shorthand}) was performed with
Monte Carlo methods. The Coulomb plane waves are solution to Schr\"odinger's
equation for a wave where the outgoing wave has momentum ${\bf q}$.
\begin{equation}
-\nabla^2 \psi({\bf q},{\bf r})+q\frac{2\eta}{r}\psi({\bf q},{\bf r})
=q^2 \psi({\bf q},{\bf r}),
\end{equation}
where $\eta$ is the Sommerfeld parameter,
\begin{equation}
\eta=\frac{Z_aZ_b \mu e^2}{q},~~\mu=\frac{E'_aE'_b}{(E'_a+E'_b)}.
\end{equation}
The usual definition of the reduced mass $\mu$, involving $m_a$ and $m_b$, has
been altered to achieve consistency with relativistic treatments for small
$e^2$.

In order to understand the form of the squared wave function, it it insightful
to compare to the classical approximation where the squared wave function in
Eq. (\ref{eq:corrdef_shorthand}) is replaced by the ratio of the initial and
final phase space.
\begin{equation}
\label{eq:classical}
\begin{split}
|\psi({\bf q},{\bf r})|^2&\rightarrow \frac{d^3q_0}{d^3q}
=\frac{1+\cos\theta_{qr}-\eta/(qr)}
{\sqrt{(1+\cos\theta_{qr})^2-2(1+\cos\theta_{qr})\eta/(qr)}}\\
&\hspace*{50pt}\cdot\Theta(1+\cos\theta_{qr}-2\eta/(qr)),
\end{split}
\end{equation}
where $\theta_{qr}$ is the angle between ${\bf q}$ and ${\bf r}$. Integrating
over $\cos\theta_{qr}$, one finds the angle-averaged correlation weight,
\begin{equation}
\frac{q_0^2dq_0}{q^2dq}
=\sqrt{1-\frac{2\eta}{qr}}\approx 1-\frac{\eta}{qr}.
\end{equation}
In the non-relativistic limit this approaches unity as $1/q^2$. Convoluting
this expression with the Gaussian source function gives a classical result for
the correlation function which is remarkably close to the quantum correlation
function for large $q$ as can be seen in Fig. \ref{fig:corr_pk}. Thus, the
tail of the correlation function provides a measure of the expectation,
$\langle 1/r\rangle$.

Extracting the shape of the source requires measuring the correlation function
as a function of the direction of ${\bf q}$. Since the correlation approaches
unity as $1/q^2$, we recommended plotting $q^2[C(q)-1]$ rather than $C(q)$ so
that the main $q$ dependence can be ignored, allowing the use of larger bins in
$q$.

\begin{figure}[t]
\centerline{\includegraphics[width=0.4\textwidth]{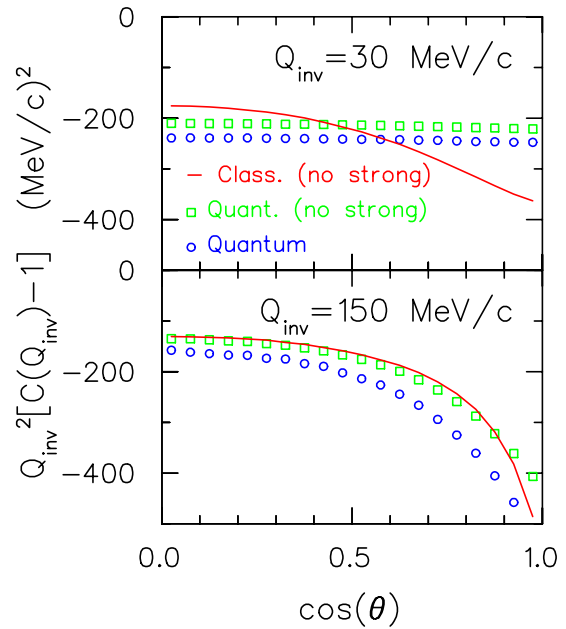}}
\caption{\label{fig:ctheta_pk} $pK^+$ correlations are shown as a function of
the angle of the relative momentum relative to the outwards direction for the
Gaussian source ($R_{\rm long}=R_{\rm side}=4$ fm, $R_{\rm out}=8$ fm). The
classical approximation becomes reasonable for large $Q_{\rm inv}$ where the
ratio of the suppression at $\cos\theta=0$ to the suppression at $\cos\theta=0$
approaches $(R_{\rm out}/R_{\rm side})^2$.}
\end{figure}
As a function of the direction of ${\bf q}$, correlations are shown in
Fig. \ref{fig:ctheta_pk} for $Q_{\rm inv}=30$ and 150 MeV/c alongside the
analogous results calculated with the classical form described in
Eq. (\ref{eq:classical}). The negative correlation at $\cos\theta=\pm 1$
derives from the fact that the relative momentum of the two positive particles
are deflected away from the direction defined by their relative position by the
repulsive Coulomb force. Careful analysis of the classical expression,
Eq. (\ref{eq:classical}), shows that for small $\eta$, the negative correlation
is confined to ${\bf q}$ and ${\bf r}$ being anti-parallel. Since the chance of
the particles being separated along the outwards axis is larger by a factor of
$R_{\rm out}^3/R_{\rm \perp}^3$, and since the strength of the correlation in
the two directions scales as $R_{\perp}/R_{\rm out}$, the strength of the
Coulomb induced correlation will be stronger along the $R_{\rm out}$ direction
by a factor of $R_{\rm out}^2/R_{\rm side}^2$ at large $q$.

It should not be particularly difficult to obtain the required statistics even
though the correlation is of the order of 1\%. Although the correlation falls
as $1/Q_{\rm inv}^2$, phase space rises as $Q_{\rm inv}^2$. Unlike
identical-particle correlations at small relative momentum, there are no issues
with two-track resolution. However, one has to consider large-scale
correlations, e.g., collective flow, jets and charge conservation.  Correlation
from charge conservation can be neglected by considering only same-sign
pairs. Collective flow can be eliminated by carefully constructing the
correlation denominator with pairs from events with the same reaction plane.
Such competing correlations ultimately limit the range in $Q_{\rm inv}$ that is
useful for the the analysis. For instance, if the uncertainty of the competing
correlations for $pK^+$ is of the order of 1\%, the analysis should be
restricted to $Q_{\rm inv}<150$ MeV/c so that Coulomb remains the dominant
factor.

The angular sensitivity of Coulomb-induced correlations have been investigated
in intermediate-energy heavy-ion collisions. These studies involved light
fragments, e.g. Carbon-Carbon, which can be treated with classical trajectories
\cite{kim}. The Coulomb mean field from the remainder of the source was found
to distort the the shape information. However, the residual Coulomb interaction
is expected to play a much smaller role at RHIC where the hadrons move much
faster and spend less time interacting with the mean field. Furthermore,
analyses can be performed with both positive and negative pairs, e.g., both
$pK^+$ and $\bar{p}K^-$. Averaging the two correlation functions should largely
cancel the effects of the residual interaction. At high energy, the angular
dependence of Coulomb correlations of non-identical particles have been used to
determine the degree to which one species is displaced relative to another.
These studies involved comparing correlations for $\theta<\pi/2$ with
correlations with $\theta>\pi/2$ \cite{lednicky,panitkin}.

Incorporating the strong interaction into the wave function in
Eq. (\ref{eq:corrdef_shorthand}) is straight-forward if the the strong
interaction can be expressed as a non-relativistic potential. One must simply
solve the Schr\"odinger equation. However, many interactions can not be
expressed in terms of such potentials. For instance, the interaction of a
$\pi^+$ and proton through the delta resonance involves a quantum rearrangement
of the participating quarks. Thus, a wave function is not a meaningful quantity
for separations below $\epsilon \sim 1$ fm, but it is certainly a well defined
object for $r>\epsilon$, and can be expressed in terms of a Coulomb wave with
the incoming partial waves are modified by phase shifts.
\begin{equation}
\label{eq:partialwaveexp}
\begin{split}
\psi({\bf q},{\bf r})&=\psi_0({\bf q},{\bf r})
+\sum_\ell \sqrt{4\pi(2\ell+1)} \frac{i^\ell}{2qr} e^{-i\sigma_\ell}\\
&\hspace*{-15pt}\cdot\left(F_\ell(\eta,qr)-iG_\ell(\eta,qr)\right)
\left(e^{-2i\delta_\ell}-1\right)Y_{\ell,m=0},\\
\sigma_\ell&\equiv\arg\left(\Gamma(1+\ell+i\eta)\right).
\end{split}
\end{equation}
Here, $F_\ell$ and $G_\ell$ are the regular and irregular partial Coulomb waves
and the second term describes the distortion of the incoming partial wave,
$F_\ell-iG_\ell$. 

For the $p\pi^+$ and $pK^+$ examples discussed in this study, the plane wave
also has a spin index, $m_s=\pm 1/2$, which is not conserved for partial waves
with $\ell>0$. The states of a given $m_s$ must be decomposed in terms of
eigenstates of total angular momentum which are phase shifted by eigen-phases,
$\delta_{J,\ell}$.  After applying some angular momentum algebra,
Eq. (\ref{eq:partialwaveexp}) can be modified to include flipping the spin. For
$\ell=1$,
\begin{equation}
\begin{split}
(e^{-2i\delta_\ell}-1)Y_{\ell,m=0} &\rightarrow\\
&\hspace*{-70pt}\left(\frac{2}{3}e^{-2i\delta_{\ell,J=3/2}}
       +\frac{1}{3}e^{-2i\delta_{\ell,J=1/2}}-1\right)
Y_{\ell,m=0}|\uparrow\rangle\\
&\hspace*{-50pt}+\frac{\sqrt{2}}{3}\left(
e^{-2i\delta_{\ell,J=3/2}}-e^{-2i\delta_{\ell,J=1/2}}\right)
Y_{\ell,m=1}|\downarrow\rangle
\end{split}
\end{equation}
Thus, the wave function for $r>\epsilon$ can be evaluated if given the phase
shifts.

For $r<\epsilon$, one must use an effective form for 
$|\psi({\bf q},{\bf r})|^2$. Since $\epsilon$ is much smaller than any
characteristic dimension of the source, only the integral of $\psi^2$ matters
in the region less than $\epsilon$, and one can safely choose,
\begin{equation}
|\psi({\bf q},r<\epsilon)|^2=|\psi_0({\bf q},{\bf r})|^2+
W(\epsilon,q),
\end{equation}
where $\psi_0$ is the Coulomb wave and $W$ is independent of $r$.
The change in the density of states can be expressed both in terms of phase
shifts and wave functions \cite{boal},
\begin{eqnarray}
\label{eq:dndqconstraint}
\Delta \frac{dN}{dq}&=&\frac{4\pi q^2}{(2\pi)^3}\int d^3r\left(
\left|\phi({\bf q},{\bf r})\right|^2-\left|\phi_0({\bf q},{\bf r})\right|^2
\right)\\
\nonumber
&=&\sum_\ell\frac{(2\ell+1)}{\pi}\frac{d\delta_\ell}{dq}\\
\nonumber
&=&\frac{2q^2\epsilon^3}{3\pi}W(\epsilon,q)\\
\nonumber
&&\hspace*{-30pt}+\sum_\ell\frac{(2\ell+1)}{2\pi}
\int_\epsilon^\infty  dr \left(\left| \phi_\ell(\eta,qr)\right|^2
-|F_\ell^2(\eta,qr)|^2\right),\\
\nonumber
&&\hspace*{-30pt}\phi_\ell=F_\ell+\frac{1}{2}(e^{-2i\delta_\ell}-1)
\left(F_\ell-iG_\ell\right).
\end{eqnarray}
Thus, $W$ can be expressed in terms of derivatives of the phase shifts and
integrals of the type,
\begin{equation}
I_\ell(\epsilon,q,\delta_\ell)\equiv \int_\epsilon^\infty
dr |\phi_\ell(\eta,qr,\delta_\ell)|^2,
\end{equation}
Since $\phi_\ell$ is a solution to the Schr\"odinger equation, one can rewrite
$I_\ell$, assuming that $\phi$ and $\phi'$ are solutions with eigenvalues $q$
and $q'\sim q$,
\begin{equation}
\begin{split}
(q^{\prime 2}-q^2)I_\ell(\epsilon,q,\delta_\ell)&=\Re\int_\epsilon^\infty dr 
\left(\partial_r^2\phi_\ell'^*\phi_\ell
-\phi_\ell^{\prime *}\partial_r^2\phi_\ell\right)\\
I_\ell(\epsilon,q,\delta_\ell)&=\left.\frac{1}{2q}\Re
\left(\partial_r\phi_\ell^*\partial_q\phi_\ell
-\phi_\ell^*\partial_r\partial_q\phi_\ell\right)\right|_\epsilon^\infty.
\end{split}
\end{equation}
Transforming the derivatives to the variables $\eta$ and $x=qr$ facilitates
use of the recursion relations for the Coulomb wave functions
\cite{abramowitzstegun},
\begin{equation}
\begin{split}
I_{\ell\ne 0}(\epsilon,q,\delta_\ell)
&=\left(\phi_\ell^*\phi_\ell+\phi_{\ell-1}^*\phi_{\ell-1}\right)
\frac{x}{2q\ell^2}(\ell^2+\eta^2)\\
&\hspace*{-40pt}
-\Re(\phi_{\ell-1}^*\phi_\ell)\frac{(2\ell+1)(\ell^2+\eta^2)\ell+2\eta
  x(\ell^2+\eta^2)+\ell\eta^2}
{2q\ell^2\sqrt{\ell^2+\eta^2}}\\
&\hspace*{-40pt}
+\left.\Re\left(\phi_{\ell}\partial_\eta 
\phi_{\ell-1}^*-\phi_{\ell-1}\partial_\eta \phi_\ell\right)
\frac{\eta\sqrt{\ell^2+\eta^2}}{2q\ell}\right|_{x=q\epsilon},\\
I_{\ell=0}(\epsilon,q,\delta_\ell)&=(\phi^*_0\phi_0+\phi^*_1\phi_1)
\frac{x(1+\eta^2)}{2q}\\
&-\Re(\phi^*_0\phi_1)\frac{(1+\eta^2)(1+2\eta x)+\eta^2)}{2q\sqrt{1+\eta^2}}\\
&+\left.\Re(\phi^*_1\partial_\eta\phi_0-\phi^*_0\partial_\eta\phi_1)
\eta\frac{\sqrt{1+\eta^2}}{2q}\right|_{x=q\epsilon}.
\end{split}
\end{equation}
The upper limit, $x=\infty$, need not be evaluated since it would be canceled
by an equal and opposite contribution from the non-phaseshifted term in
Eq. (\ref{eq:dndqconstraint}). For the case with no Coulomb interactions, one
can derive a simpler form involving spherical Bessel functions rather than
Coulomb wave functions. For the no-Coulomb example, one can show that
$W=0$ as $q\rightarrow 0$.

\begin{figure}[t]
\centerline{\includegraphics[width=0.4\textwidth]{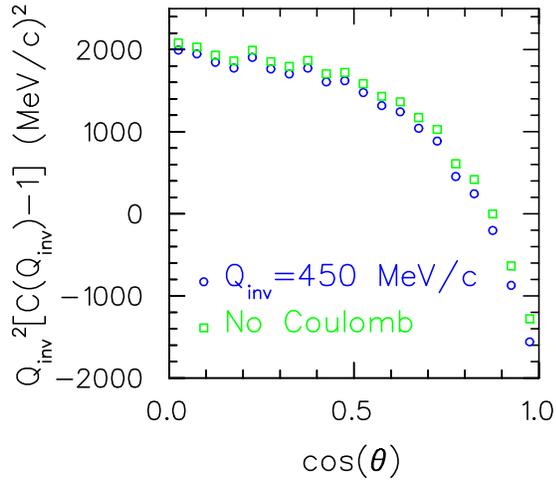}}
\caption{\label{fig:ctheta_ppi} $p\pi^+$ angular correlations for a Gaussian
  source ($R_{\rm out}=R_{\rm side}=4$ fm, $R_{\rm long}=8$ fm) with $Q_{\rm
  inv}=450$ MeV/c. At this $Q_{\rm inv}$ the correlation is dominated by the
  $\Delta^{++}$ resonance. The suppression for $\cos\theta=\pm
  1$ can be explained by shadowing.}
\end{figure}
Using experimentally tabulated phase shifts, and the arbitrary choice of
$\epsilon=1$ fm, wave functions were numerically generated and convoluted with
the source function through Eq. (\ref{eq:corrdef_shorthand}) to generate
correlation functions. Correlations for $pK^+$ were only slightly affected by
the strong interaction as can be seen in Fig.s \ref{fig:corr_pk} and
\ref{fig:ctheta_pk}. For this example, there are no resonant channels and the
correlation is dominated by Coulomb. Figure \ref{fig:ctheta_ppi} shows results
for $p\pi^+$ correlations which are dominated by the $\Delta^{++}$ for $Q_{\rm
inv}$ near the resonant momentum, $Q_{\rm inv}=450$ MeV. A strong angular
correlation is caused by shadowing in the forward direction. Like
Coulomb-induced correlation, there is a dip when the relative momentum is
aligned with the long-axis of the source.

Strong and Coulomb induced correlations had been previously studied for their
ability to to unfold the angle-averaged source function, $g(r)$, for both
high-energy and low-energy collisions
\cite{browndanielewicz,e895imaging,verdeimaging}. Our findings show that such
correlations also have tremendous potential to discern shape characteristics
and thus provide an estimate of source lifetimes. Determining shape and
lifetime characteristics had been previously confined to analyses of
identical-particle correlations.  As strong and Coulomb correlations are of a
manifestly different character than correlations from identical-particle
interference, the analyses described here represent a truly independent
strategy for determining space-time characteristics of hadronic sources.

\begin{acknowledgments}
This work was supported by the National Science Foundation, Grant No.
PHY-00-70818. S. Petriconi gratefully acknowledges the support of the
Studien\-stiftung Foundation. The authors also thank D.A. Brown for very
helpful comments.
\end{acknowledgments}


\end{document}